\documentclass[preprintnumbers,amsmath,amssymb]{revtex4}
\usepackage{mathrsfs}%====================================
\usepackage{amssymb}
\usepackage{graphicx}% Include figure files
\usepackage{dcolumn}% Align table columns on decimal point
\usepackage{bm}% bold math
\usepackage{textcomp}
\usepackage{amsmath}
\usepackage[titletoc]{appendix}

\newcommand\ket[1]{\ensuremath{|#1\rangle}}

\newcommand\iprod[2]{\ensuremath{\langle#1|#2\rangle}}
\newcommand\oprod[2]{\ensuremath{|#1\rangle\langle#2|}}

\newcounter{RomanNumber}
\newcommand{\MyRoman}[1]{\setcounter{RomanNumber}{#1}\Roman{RomanNumber}}

\begin{document}
\title{Round robin differential phase shift quantum key distribution with yes-no detectors only}
\author{ Cong Jiang$ ^{1,2}$, Zong-Wen Yu$ ^{1,3}$,
and Xiang-Bin Wang$ ^{1,2,4,5\footnote{Email Address: xbwang@mail.tsinghua.edu.cn}\footnote{Also at Center for Atomic and Molecular Nanosciences, Tsinghua University, Beijing 100084, China}}$}

\affiliation{ \centerline{$^{1}$State Key Laboratory of Low
Dimensional Quantum Physics, Department of Physics,} \centerline{Tsinghua University, Beijing 100084,
Peoples Republic of China}
\centerline{$^{2}$ Synergetic Innovation Center of Quantum Information and Quantum Physics, University of Science and Technology of China}\centerline{  Hefei, Anhui 230026, China
 }
\centerline{$^{3}$Data Communication Science and Technology Research Institute, Beijing 100191, China}\centerline{$^{4}$ Jinan Institute of Quantum technology, SAICT, Jinan 250101,
Peoples Republic of China}
\centerline{$^{5}$Department of Physics, Southern University of Science and Technology, Shenzhen, 518055, People’s Republic of China.}
}
%%%%%%%%%%%%%%%%%%%%%%%%%%%%%%%%%%%%%%%%%%%%%%%%%%%%%%%%%%%%%%%%%%%
%%%%%%%%%%%%%%%%%%%%%%%%%%%%%%%%%%%%%%%%%%%%%%%%%%%%%%%%%%%%%%%%%%%
%%%%%%%%%%%%%%%%%%%%%%%%% Abstract %%%%%%%%%%%%%%%%%%%%%%%%%%%%%%%%
\begin{abstract}
In the original round-robin differential-phase-shift (RRDPS) quantum key distribution and its improved method, the photon-number-resolving detectors are must for the security. We present a RRDPS protocol with yes-no detectors only. We get the upper bounds of mutual information of Alice and Eve, and Bob and Eve, and the formula of key rate. Our main idea is to divide all counts into two classes, the counts due to the odd number of photons incident to the detectors and the counts due to the even number photons incident to the detectors. The fact that the bit-flip error rate of the later class is certainly $50\%$ makes it possible for us to perform a tightened estimation of the upper bound of the leakage information. The robustness of original RRDPS against source flaws such as side-channel attacks still holds for the RRDPS with yes-no detectors. The simulation results show that the key rate of RRDPS with yes-no detectors is close to that of RRDPS with photon-number-resolving detectors. Our results make the RRDPS protocol much more practical.
\end{abstract}

%%%%%%%%%%%%%%%%%%%%%%%%%%%%%%%%%%%%%%%%%%%%%%%%%%%%%%%%%%%%%%%%%%%
%%%%%%%%%%%%%%%%%%%%%%%%%%%%%%%%%%%%%%%%%%%%%%%%%%%%%%%%%%%%%%%%%%%
%%%%%%%%%%%%%%%%%%%%%%%%%%%%%%%%%%%%%%%%%%%%%%%%%%%%%%%%%%%%%%%%%%%

\maketitle

\noindent\textit{Introduction} Quantum key distribution (QKD) provides a information-theoretical secure key distribution method between two parties, Alice and Bob, and the security is based on the laws of quantum physics. Since Bennett and Brassard proposed the first QKD protocol~\cite{bennett1984quantum}, BB84 protocol, many new QKD protocols~\cite{braunstein2012side,lo2012measurement,grosshans2002continuous,lu2018overcoming,tamaki2018information,wang2018twin,
ma2018phase,cui2019twin,curty2018simple,jiang2019unconditional,xu2019general} and improved methods~\cite{hwang2003quantum,wang2005beating,lo2005decoy} have been proposed to assure the security with imperfect devices and improve the key rate. The key of those QKD's security is that any eavesdropping of Eve would introduce disturbances, and we could evaluate the upper bound of leakage information according the disturbances. Surprisingly, a QKD protocol named round robin differential phase shift ((RRDPS)) QKD protocol~\cite{sasaki2014practical} was proposed in which the leakage information could be evaluated without monitoring any signal disturbance. This is so remarkable that has attracted many attentions~\cite{guan2015experimental,takesue2015experimental,wang2015experimental,mizutani2015robustness,
yin2018improved,matsuura2019refined}. Not only has it greatly simplified to realization in practice because there is no bases-switching, but also improved the security in side-channel aspect~\cite{mizutani2015robustness}. Notabaly, Yin et al.~\cite{yin2018improved} proposed a tighter upper bound of leakage information and greatly improved the key rate especially with small $L$ compared with the original RRDPS. 

The detectors used in original RRDPS~\cite{sasaki2014practical,yin2018improved,matsuura2019refined} are photon-number-resolving (PNR) detectors.This limits the practical application of RRDPS. Thus it is crucially important to study the RRDPS protocol with yes-no detectors only. Here we propose such a RRDPS protocol. Our main idea is to divide all counts into two classes, the counts due to the odd number photons incident to the detectors and the counts due to the even number photons incident to the detectors. The fact that the bit-flip error rate of the later class is certainly $50\%$ makes it possible for us to perform a tightened estimation of the upper bound of the leakage information.

This paper is arranged as follows. We first introduce the original RRDPS protocol. We then present the main results of the upper bound of leakage information and the formula of key rate of RRDPS with yes-no detectors. After that, we present some numerical results of RRDPS with yes-no detectors and compare with RRDPS with PNR detectors. The article ends with some concluding remarks.

\begin{figure}
\centering
\includegraphics[width=9cm]{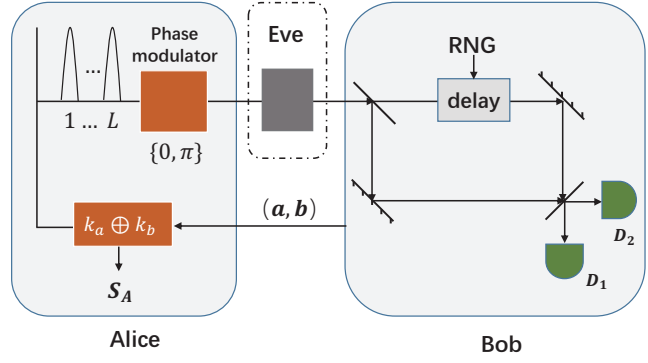}
\caption{Alice first prepares a train of $L$ pulses and randomly modulates their phases into $0$ or $\pi$ and sends this pulses to Bob. Bob performs interference measurement to the incoming pulses with the set-up above. In this set-up, the two detectors are PNR detectors and the delay module will randomly produce $r(r\in[1,L-1])$ delay. If only one detector response one time in the whole detection process, we call this $L$-pulses train caused a count and Alice and Bob will record the corresponding bit as the sifted key.}\label{model}
\end{figure}

\noindent\textit{Main Results} As shown in Figure.~\ref{model}, in the original RRDPS protocol~\cite{sasaki2014practical}, the communication parties are Alice and Bob. Alice first prepares a train of $L$ pulses and randomly modulates their phases into $0$ or $\pi$. If Alice's source is single-photon source, she will finally prepare the state as
\begin{equation}\label{singlestate}
\ket{\Psi}_A=\frac{1}{\sqrt{L}}\sum_{i=1}^L(-1)^{k_i}\ket{i},
\end{equation}   
where $\ket{i}$ denotes there is one photon in the $i$th pulse and $k_i$ is randomly $1$ or $0$. If Alice's source is weak coherent state source, the final state she prepares is
\begin{equation}
\ket{\Psi}_{A}^{'}=\otimes_{i=1}^L\ket{(-1)^{k_i}\alpha},
\end{equation} 
where $\ket{\alpha}$ denotes there is a weak coherent state pulse with intensity $|\alpha|^2$ in location $i$. Then Alice sends this $L$-pulses train to Bob, and Bob performs interference measurement to the incoming pulses with the set-up shown in Figure.~\ref{model}. In this set-up, the two detectors are PNR detectors and the delay module will randomly produce $r(r\in[1,L-1])$ delay. If only one detector responses one time in the whole detection process, we call this $L$-pulses train causes a count and Bob will announce the detection result $\{a,b\},b=a+r(mod\quad L)$ to Alice through a public channel. Bob records the measured phase difference as his sifted key bit $S_B$ and Alice records $S_A=k_a\oplus k_b$ as her sifted key bit. 

In the original RRDPS protocol, to evaluate the maximum leakage information, the PNR detectors are needed~\cite{sasaki2014practical,yin2018improved,matsuura2019refined}, or else the formula of key rate will no longer hold. But as so far, the technology of PNR detectors are difficult. Here we study the maximum leakage information and key rate formula if the two detectors are yes-no detectors in Figure.~\ref{model}. We take the single photon case as an example to briefly introduce our conclusions, and the other cases and detailed proof is shown in Supplementary materials. 

Inspired by Ref.~\cite{yin2018improved}, Eve's optimal collective attack can be given by the following equation if Alice prepares her state as Eq.~\eqref{singlestate}:
\begin{equation}
U_{eve}\ket{i}\ket{e_{000}}=\sum_{n=0}^\infty\sum_{j=1}^L c_{ijn}\ket{n_j}\ket{e_{ijn}},
\end{equation}
where $\ket{e_{000}}$ is the initial state of Eve's ancilla bits; $\ket{e_{ijn}}$ is the final state of Eve's ancilla state after attacking and $\ket{n_j}$ represents there are $n$ photons in the $j$th pulse. The coefficient $c_{ijn}$ satisfies $\sum_{n=0}^\infty\sum_{j=1}^L |c_{ijn}|^2=1$. While considering the leakage information, only the states of $n=1$ work if the detectors are PNR detectors, and this is just the case of Ref.~\cite{yin2018improved}. But if the detectors are yes-no detectors, we have to take all the $n>0$ states into consideration. 

Consider the mutual information of Alice and Eve, $I_{AE}$, we have that the state of Alice and Eve after Eve's attacking is
\begin{equation}
\ket{\Psi}_{AE}= \sum_{n=0}^\infty\sum_{i=1}^L\sum_{j=1}^L (-1)^{k_i}c_{ijn}\ket{n_j}\ket{e_{ijn}}.
\end{equation}

If Bob measures such an incoming state and announces $\{a,b\}$, the density matrix (non-normallized) of Eve's ancilla bits will be~\cite{notation}
\begin{equation}\label{evestate}
\begin{split}
\rho_E=&\frac{1}{2}\sum_{n=1}\{P[\sum_{i}(-1)^{k_i}\frac{\widetilde{c}_{ian}}{2^n}+\sum_{i}(-1)^{k_i}\frac{\widetilde{c}_{ibn}}{2^n}]\\
&+P[\sum_{i}(-1)^{k_i}\frac{\widetilde{c}_{ian}}{2^n}+\sum_{i}(-1)^{k_i}(-1)^n\frac{\widetilde{c}_{ibn}}{2^n}]\},
\end{split}
\end{equation}
where $P(\ket{x})\equiv \oprod{x}{x}$ and $\widetilde{c}_{ijn}\equiv c_{ijn}\ket{e_{ijn}}$. The first part of Eq.~\eqref{evestate} is caused by detector $D_1$, and the second part is caused by detector $D_2$. We find that the upper bounds of $I_{AE}$ corresponding to $n=1,3,5,\dots$ and $n=2,4,6,\dots$ are totally different. Here we call the counting events of $n=1,3,5,\dots$ as odd-counts and the counting events of $n=2,4,6,\dots$ as even-counts. 

If the single photon pulses train causes an odd-count, the density matrix of Eve is
\begin{equation}
\rho_{E}^{Odd}=\sum_{n=Odd}\{P[\sum_{i}\frac{(-1)^{k_i}\widetilde{c}_{ian}}{2^n}]+P[\sum_{i}\frac{(-1)^{k_i}\widetilde{c}_{ibn}}{2^n}]\}.
\end{equation}
The value of $k_i(i\neq a,b)$ is randomly $0$ or $1$, thus we can average $\rho_{E}^{Odd}$ with random $k_i(i\neq a,b)$ to simplify the calculation. Without compromising the security, we can assume
\begin{equation}\label{assum1}
\iprod{e_{ijn}}{e_{klm}}=\delta_{ik}\delta_{jl}\delta_{nm}.
\end{equation}
Finally only the states $P[\frac{\widetilde{c}_{aan}}{2^n}\pm\frac{\widetilde{c}_{ban}}{2^n}]$ and $P[\frac{\widetilde{c}_{abn}}{2^n}\pm\frac{\widetilde{c}_{bbn}}{2^n}]$ accounts for the Holevo bound which measures the maximum mutual information of two parties if they share a system in quantum information. Thus we can get the upper bound of $I_{AE}$
\begin{equation}\label{iaeodd}
I_{AE}^{Odd}\le\phi(1,L)=Max_{x_1,x_2}\frac{\varphi((L-1)x_1,x_2)}{L-1},
\end{equation}
where $\varphi(x,y)=-x\log_2{x}-y\log_2{y}+(x+y)\log_2{(x+y)}$ and $x_1+x_2=1$. Note that the assumption of Eq.~\eqref{assum1} would certainly introduce $50\%$ bit-flip error rate, thus Eq.~\eqref{iaeodd} shows that the maximum leakage information of odd-counts is limited even if Eve's attack is optimal and introduces $50\%$ bit-flip error rate. And in general case, $\phi(1,L)<H(\frac{1}{L-1})$, where $H(x)$ is Shannon entropy, $H(x)=-x\log_2{x}-(1-x)\log_2{(1-x)}$, and $H(\frac{1}{L-1})$ is the upper bound of leakage information proposed in original RRDPS~\cite{sasaki2014practical}. 

In Supplementary materials, we prove that if the single photon pulses train causes an even-count, the upper bound of $I_{AE}^{Even}$ equals $1$. Dose this mean that we can not extract any secure final keys if we don't know whether the counts is an odd-count or even-count? Luckily, the answer is no. We next consider the mutual information of Bob and Eve, $I_{BE}$. If the count $\{a,b\}$ is detected by $D_1$, the density matrix of Eve is
\begin{equation}\label{eq10}
\rho_E^{D_1}=\sum_{n=1}P[\sum_{i}(-1)^{k_i}\frac{\widetilde{c}_{ian}}{2^n}+\sum_{i}(-1)^{k_i}\frac{\widetilde{c}_{ibn}}{2^n}].
\end{equation} 
If the count $\{a,b\}$ is detected by $D_2$, the density matrix of Eve is
\begin{equation}\label{eq11}
\rho_E^{D_2}=\sum_{n=1}P[\sum_{i}(-1)^{k_i}\frac{\widetilde{c}_{ian}}{2^n}+\sum_{i}(-1)^{k_i}(-1)^n\frac{\widetilde{c}_{ibn}}{2^n}].
\end{equation} 

From Eqs.~\eqref{eq10} and \eqref{eq11}, it is easy to see that $\rho_E^{D_1}$ and $\rho_E^{D_2}$ are the same if $n$ are even numbers, which means Eve can not distinguish whether an even-count is caused by $D_1$ or $D_2$ at all, thus $I_{BE}^{Even}=0$. Different from the situation of $I_{AE}^{Odd}$, this result needs not the average of random phase and orthogonality assumption Eq.~\eqref{assum1}. If the single photon cause an odd-count, the upper bound of $I_{BE}^{Odd}$ equals $1$. The detailed proof of single photon case and other general cases are shown in Supplementary materials. 

We list the main results as following:\\ 
\textit{If Alice sends out a train of $L$ pulses contained $N(N\le \frac{L}{2})$ photons, and Bob announces this pulses train causes a count, the upper bounds of mutual information of Alice and Eve, $I_{AE}$, and the mutual information of Bob and Eve, $I_{BE}$, are}
\begin{equation}\label{conclusion111}
\begin{split}
I_{AE}^{Odd}(N)\le \phi(N,L),\quad I_{BE}^{Even}(N)=0,
\end{split}
\end{equation} 
\textit{where}
\begin{equation}
\phi(N,L)=\max\limits_{x_1,x_2,\dots,x_{N+1}}\left\{\frac{\sum\limits_{k=1}^N\varphi((L-k)x_k,kx_{k+1})}{L-1}\right\},
\end{equation}
\textit{and $\sum_{k=1}^{N+1}x_k=1$. And $I_{AE}^{Even}=1,I_{BE}^{odd}=1$.}

In short, \textbf{\MyRoman{1}}. If the train causes an odd-count, the maximum mutual information of Alice and Eve is $\phi(N,L)$, but the maximum mutual information of Bob and Eve is $1$. \textbf{\MyRoman{2}}. If the train causes an even-count, the maximum mutual information of Bob and Eve is $0$, but the maximum mutual information of Alice and Eve is $1$. This means if the final counts contain many odd-counts, the mutual information of Alice and Eve is relatively small while the mutual information of Bob and Eve is pretty large, and vice versa. Thus if the error correction step is based on Alice's or Bob's sifted raw key bits, the final key bits we extract must be zero, since we have no idea about the ratio of odd-counts or even-counts. But if the sifted raw key bits are randomly split into two parts, the first part accounts for $\gamma$, and the second part accounts for $1-\gamma$. The error correction of the first part is based on Alice's sifted raw key bits and the second part is based on Bob's sifted raw key bits. The final key rate is
\begin{equation}\label{keyrate}
\begin{split}
LR=\max_{\gamma} \{&\gamma \max[I_{AB}-I_{AE},0]\\
&+(1-\gamma)\max[I_{AB}-I_{BE},0]\}.
\end{split}
\end{equation}
And in general case, at least one of $I_{AB}-I_{AE}$ and $I_{AB}-I_{BE}$ is greater than zero.
    
If Alice's source is weak coherent state source, and she randomly modulates the mutual phase the different pulses in the train, the state that she prepares is equivalent to the classical mixture of different photon numbers. And the density matrix of the pulses train with intensity $\mu$ in photon number space is 
\begin{equation}
\rho(\mu)=\sum_{k=0}a_k\oprod{k}{k},\quad a_k=\frac{\mu^ke^{-\mu}}{k!}.
\end{equation}
 
We denote the total counting rate and error rate of the $L$-pulses train are $Q$ and $E$, and the counting rate and error rate of $k$-photons $L$-pulses train are $Y_k$ and $e_k$, in which the ratio of odd-counts is $\alpha_k$ and even-counts is $\beta_k=1-\alpha_k$, we have
\begin{equation}\label{iaeup}
\begin{split}
I_{AE}\le&\sum_{n=1}a_nY_n[\alpha_n\phi(n,L)+\beta_n]\\
=&\sum_{n=1}^{n_{th}}a_nY_n[\alpha_n\phi(n,L)+\beta_n]+\sum_{n>n_{th}}a_nY_n\\
\le & \alpha(Q-e_{src})\phi(n_{th},L)+\beta(Q-e_{src})+e_{src},
\end{split}
\end{equation}
where $e_{src}=\sum_{n>n_{th}}a_n$ and $\alpha+\beta=1$. We use a fact in the last step of Eq.~\eqref{iaeup}
\begin{equation}
\sum_{n=0}^{n_{th}} a_nY_n\alpha_n+\sum_{n=0} ^{n_{th}}a_nY_n\beta_n=\sum_{n=0} ^{n_{th}}a_nY_n\ge Q-e_{src}. \nonumber
\end{equation} 
And similarly we have
\begin{equation}\label{ibeup}
\begin{split}
I_{BE}\le\alpha(Q-e_{src})+e_{src}.
\end{split}
\end{equation}
Here the $n_{th}$ is the boundary of tagged photons ($n>n_{th}$) and untagged photons ($n\ge n_{th}$) and will be an optimized parameters in the simulation parts to get a better key rate.

The Eqs.~\eqref{eq10} and \eqref{eq11} show us an important fact that if a single photon count is an even-count, its bit-flip error rate definitely is $1/2$, whether Eve attacks it or not. This is different from the case of odd-counts, where the bit-flip error rate is $1/2$ only if Eve attacks all the key bits with his optimal attack strategy. And we show that this fact holds for any $k-$photon counts in Supplementary materials. In fact, we can assume Eve's attacks introduce $e_k^{odd}$ bit-flip error rate of the odd-counts, and we have
\begin{equation}
\alpha_k e_k^{odd}+\frac{1}{2}\beta_k=e_k,
\end{equation}
which implies $\beta_k\le 2e_k$. Thus we have $\beta(Q-e_{src})\le 2EQ$, which implies 
\begin{equation}\label{alpha}
\alpha\ge 1-\frac{2EQ}{Q-e_{src}},
\end{equation}
if $Q-e_{src}>0$.  

Combine Eqs.~\eqref{keyrate}-\eqref{ibeup} and we could get the final key rate with observable values of experiment. Finally, we have the following formulas of final key rate
\begin{equation}\label{finalkeyrate}
\begin{split}
&LR=\max\limits_{\gamma}\min\limits_{\alpha}\{\gamma \max[R_1(\alpha),0]+(1-\gamma)\max[R_2(\alpha),0]\},\\
&s.t. \quad 1-\frac{2EQ}{Q-e_{src}}\le \alpha \le 1,\quad 0\le\gamma\le1 \\
&R_1(\alpha)=\alpha(Q-e_{src})(1-\phi(n_{th},L))-fQH(E),\\
&R_2(\alpha)=(1-\alpha)(Q-e_{src})-fQH(E).
\end{split}
\end{equation}

\begin{table}%\footnotesize
\begin{ruledtabular}
\begin{tabular}{ccccc}
$\eta_d$ & $p_d$ &  $f$  & $e_0$&$\alpha_f$\\%%\rule{0pt}{0.3cm}\\
\hline
$40\%$& $1.0\times 10^{-7}$  & $1.15$  & $0.5$ & $0.2$\\ %%\rule{0pt}{0.3cm}\\
%%\hline
%%\hline
\end{tabular}
\end{ruledtabular}
\caption{List of experimental parameters used in numerical simulations. $\eta_d$: the detection efficiency of Bob's detectors; $p_d$: the dark counting rate of the Bob's detectors; $f$: the error correction inefficiency; $e_0$: error rate of the vacuum count; $\alpha_f$: the fiber loss coefficient ($dB/km$).}\label{exproperty}
\end{table}
\begin{table}
\begin{ruledtabular}
\begin{tabular}{cccccccc}
$E_d$ & $\mu$ &  $\alpha_{min}$ &  $\alpha$ & $\gamma$&  $L$  &  $n_{th}$ &$R$\\
\hline
$1.5\%$& $8.89$ & $0.97$ & $0.97$ & $1.0$  & $93$  &   $19$  & $4.12\times10^{-5}$\\
\hline
$3.0\%$& $6.02$ & $0.94$ & $0.94$ & $1.0$  & $97$  &   $15$  & $2.50\times10^{-5}$\\
\hline
$6.0\%$& $2.35$ & $0.87$ & $0.87$ & $1.0$  & $96$  &   $9$  & $7.52\times10^{-6}$\\
\hline
$10.0\%$& $0.54$ & $0.79$ & $0.79$ & $1.0$  & $97$  &   $5$  & $5.42\times10^{-7}$\\
\end{tabular}
\end{ruledtabular}
\caption{List of optimal parameters and the corresponding key rate in different miallignment-error probability, $E_d$. Here, we set the distance between Alice and Bob is 100 $km$, and $\mu$: the intensity of the phase-randomized weak coherent state pulses train; $\alpha_{min}$: the min value of $\alpha$ constrained by Eq.~\eqref{alpha}; $\alpha_{best}$: the optimal value of $\alpha$ according to our optimized algorithm; $\gamma$: the accounts of the part that the error correction is based on Alice's data, and defined in Eq.~\eqref{keyrate}; $L$: the length of the pulses train; $n_{th}$: the boundary of tagged and untagged photons; $R$: the final key rate.}\label{opti}
\end{table}

\begin{figure}
\centering
\includegraphics[width=8cm]{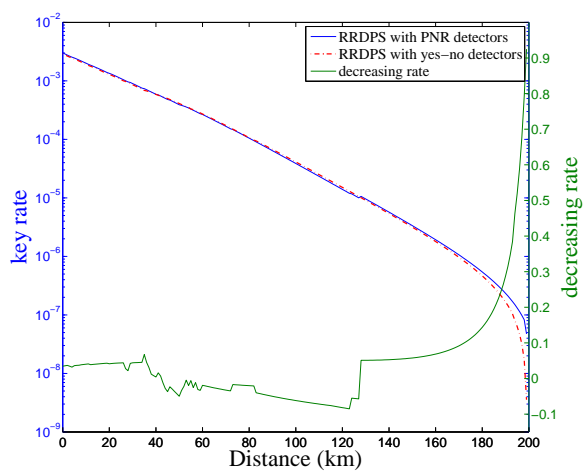}
\caption{The key rates of the original RRDPS with PNR detectors and our protocol with yes-no detectors versus the distance between Alice and Bob. The green line is the decent degree of the RRDPS with yes-no detectors compared with the key rate of RRDPS with PNR detectors. Here we set the misalignment-error probability, $E_d$, as $1.5\%$.}\label{fig1}
\end{figure}

\noindent\textit{Numerical Simulation} We then show some numerical results and compare with the results of RRDPS with PNR detectors~\cite{yin2018improved}. To clearly show the advantage of our method, we assume the properties of Bob's yes-no detectors and PNR detectors including the repetitive rate, detection efficiency and dark counting rate, are the same. Besides, we assume the PNR detectors could discriminate  the single photon from zero, two or more photons perfectly. The performance parameters of the detectors and other experiment devices are list in Table.~\ref{exproperty}. We use the linear model to simulate the observed values of the counting rate, $Q$, and error rate, $E$, of the $L$-pulses train, and the detailed of our numerical simulation method are shown in the Supplementary materials. 

We list some optimal parameters and their corresponding key rate under different misalignment-error probability, $E_d$, in Table.~\ref{opti}. Here we set the distance between Alice and Bob is 100 $km$. The data in Table.~\ref{opti} show that the while $E_d=1.5\%,3.0\%,6.0\%,10.0\%$, optimal $\gamma=1$ and $\alpha_{best}=\alpha_{min}$, where $\alpha_{min}$ is the min value of $\alpha$ constrained by Eq.~\eqref{alpha} and $\alpha_{best}$ is the optimal value of $\alpha$ according to our optimized algorithm. This implies that we can always set $\gamma=1$ and $\alpha=1-\frac{2EQ}{Q-e_{src}}$ in the key rate formula of Eq.~\eqref{finalkeyrate}. Besides, we have the following fact that:
\begin{equation*}
\begin{split}
&\max\limits_{\gamma}\min\limits_{\alpha}\{\gamma \max[R_1(\alpha),0]+(1-\gamma)\max[R_2(\alpha),0]\}\\
&\ge R_1(1-\frac{2EQ}{Q-e_{src}}),
\end{split}
\end{equation*}
where $\alpha$ is in the range constrained by Eq.~\eqref{finalkeyrate}. Thus without comprising the security, we can rewrite the formula of key rate in Eq.~\eqref{finalkeyrate} as
\begin{equation}\label{final2}
LR=[(1-2E)Q-e_{src}][1-\phi(n_{th},L)]-fQH(E).
\end{equation}
And in general case, the key rate calculated by Eq.~\eqref{final2} is the same as Eq.~\eqref{finalkeyrate} with the same experiment conditions.

Fig.~\ref{fig1} shows the key rates of the original RRDPS with PNR detectors and our protocol with yes-no detectors versus the distance between Alice and Bob, where we set the misalignment-error probability, $E_d$, as $0.015$. The blue solid line is the results of RRDPS with PNR detectors and the red dashed line is the results of RRDPS with yes-no detectors. Those two line is almost overlapped except the tail of the lines. The green solid line is the decent degree of the RRDPS with yes-no detectors compared with the key rate of RRDPS with PNR detectors, which clearly shows that their key rates differ less than $10 \% $ with the distance of Alice and Bob range in $0-175$ $km$. This results are intuitive since the formula of key rate with PNR detectors proposed in Ref.~\cite{yin2018improved} is 
\begin{equation}\label{yin}
LR_{PNR}=(Q-e_{src})[1-\phi(n_{th},L)]-fQH(E).
\end{equation}
The only difference between Eq.~\eqref{final2} and Eq.~\eqref{yin} is the coefficient, $1-2E$, of $Q$. Thus the results of RRDPS with yes-no detectors and PNR detectors are almost the same if the bit-flip error rate $E$ is small.

\noindent\textit{Conclusion} We present a protocol for RRDPS with yes-no detectors and the upper bounds of mutual information of Alice and Eve, and Bob and Eve as shown in Eq.~\eqref{conclusion111}. We present the formula of key rate of RRDPS with yes-no detectors by Eqs.~\eqref{finalkeyrate} and \eqref{final2}. Our main idea is to divide all counts into two classes, the counts due to the odd number photons incident to the detectors and the counts due to the even number photons incident to the detectors. The fact that the bit-flip error rate of the later class is certainly $50\%$ makes it possible for us to perform a tightened estimation of the upper bound of the leakage information. The major advantages of the original RRDPS protocol, the realization simplicity without bases switching and the source side-channel-free property hold in our protocol. The simulation results show that the key rate of RRDPS with yes-no detectors is close to that of RRDPS with PNR detectors.

\section{Supplementary materials}
\section{The upper bound of leakage information in the single-photon case}
\begin{figure}
\centering
\includegraphics[width=10cm]{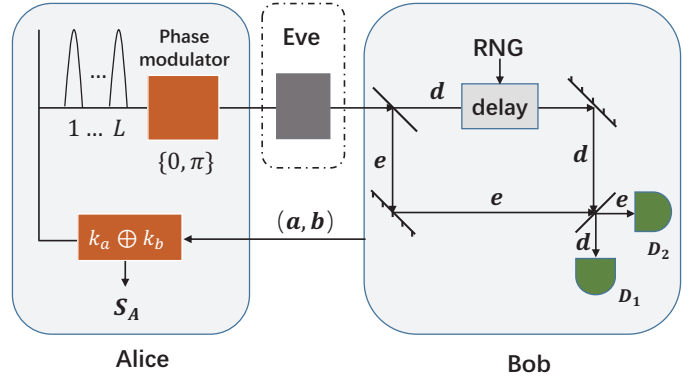}
\caption{Experiment set-ups of RRDPS. Alice first prepares a train of $L$ pulses and randomly modulates their phases. Then Alice sends this $L$-pulses train to Bob, and Bob performs interference measurement to the incoming pulses. If this $L$-pulses train causes a count, Bob will announce the location of this counts $\{a,b\}$.Bob records the measured phase difference as his sifted key bit $S_B$ and Alice records $S_A=k_a\oplus k_b$ as her sifted key bit.}\label{model}
\end{figure}

 As shown in Figure.~\ref{model}, Alice first prepares a train of $L$ pulses and randomly modulate their phases. If Alice's source is single-photon source, the state she finally prepares is
\begin{equation}\label{singlestate}
\ket{\Psi}_A=\frac{1}{\sqrt{L}}\sum_{i=1}^L(-1)^{k_i}\ket{i},
\end{equation}   
where $\ket{i}$ denotes there is one photon in the $i$th pulse and $k_i$ is randomly $1$ or $0$. 
Inspired by the method of evaluating the upper boung of leakage information in Ref.~\cite{yin2018improved}, Eve's optimal collective attack can be given by the following equation for the state shown in Eq.~\eqref{singlestate}:
\begin{equation}
U_{eve}\ket{i}\ket{e_{000}}=\sum_{n=0}^\infty\sum_{j=1}^L c_{ijn}\ket{n_j}\ket{e_{ijn}},
\end{equation}
where $\ket{e_{000}}$ is the initial state of Eve's ancilla bits; $\ket{e_{ijn}}$ is the final state of Eve's ancilla bits after attacking and $\ket{n_j}$ denotes there are $n$ photons in location $j$. The coefficient $c_{ijn}$ satisfies $\sum_{n=0}^\infty\sum_{j=1}^L |c_{ijn}|^2=1$.
After Eve's attacking, the state of Alice and Eve (unnormalized) is
\begin{equation}\label{aestate}
\ket{\Psi}_{AE}= \sum_{n=0}^\infty\sum_{i=1}^L\sum_{j=1}^L (-1)^{k_i}c_{ijn}\ket{n_j}\ket{e_{ijn}}.
\end{equation}
Through Bob's detection set-up, the state of Eq.~\eqref{aestate} is involved into 
\begin{align}
&\to \sum_{ijn}(-1)^{k_i}\widetilde{c}_{ijn}\frac{\hat{d}_j^{\dagger,n}}{\sqrt{n!}}\ket{0},\\
&\to \sum_{ijn}(-1)^{k_i}\widetilde{c}_{ijn}\frac{1}{2^n\sqrt{n!}}(\hat{d}_{j+r}^{\dagger}+\hat{e}_{j+r}^{\dagger}+\hat{d}_{j}^{\dagger}-\hat{e}_{j}^{\dagger})^n\ket{0}\equiv \ket{\Psi}_{AE}^{detect},
\end{align} 
where $\widetilde{c}_{ijn}\equiv c_{ijn}\ket{e_{ijn}}$ and $r$ is randomly chosen from $1,2,\dots,L-1$. 
\subsection{The mutual information of Alice and Eve}
We first consider the upper bound of mutual information of Alice and Eve, $I_{AE}$. If only one of Bob's detectors responses in location $k$, then only the states with superscript $j=k-r$ or $j=k$ account. We denote $a=k-r$ and $b=k$, the density matrix of Eve's ancilla bits is
\begin{equation}\label{evestate}
\begin{split}
\rho_E=\frac{1}{2}\sum_{n=1}\{P[\sum_{i}(-1)^{k_i}\frac{\widetilde{c}_{ian}}{2^n}+\sum_{i}(-1)^{k_i}\frac{\widetilde{c}_{ibn}}{2^n}]+P[\sum_{i}(-1)^{k_i}\frac{\widetilde{c}_{ian}}{2^n}+\sum_{i}(-1)^{k_i}(-1)^n\frac{\widetilde{c}_{ibn}}{2^n}]\}
\end{split}
\end{equation} 
where $P(\ket{x})\equiv \oprod{x}{x}$. The first part of Eq.~\eqref{evestate} is caused by detector $D_1$, and the second part is caused by detector $D_2$. We find that the upper bounds of $I_{AE}$ for the case $n=1,3,5,\dots$ or $n=2,4,6,\dots$ are totally different. Here we call the counting event of $n=1,3,5,\dots$ as odd-count and the counting event of $n=2,4,6,\dots$ as even-count. 
\subsubsection{The mutual information of Alice and Eve with odd-count}
If the single photon pulses train causes an odd-count, the density matrix of Eve is
\begin{equation}
\rho_{E}^{Odd}=\sum_{n=Odd}\{P[\sum_{i}\frac{(-1)^{k_i}\widetilde{c}_{ian}}{2^n}]+P[\sum_{i}\frac{(-1)^{k_i}\widetilde{c}_{ibn}}{2^n}]\}.
\end{equation}
The value of $k_i,(i\neq a,b)$ is randomly $0$ or $1$, thus we can average $\rho_{E}^{Odd}$ with random $k_i,(i\neq a,b)$ to simplify the calculation. We denote
\begin{equation}
\rho_n^{a,b}=P[\sum_{i}\frac{(-1)^{k_i}\widetilde{c}_{ian}}{2^n}]+P[\sum_{i}\frac{(-1)^{k_i}\widetilde{c}_{ibn}}{2^n}],
\end{equation}
and then we have
\begin{equation}
\begin{split}
\rho_n^{a,b}&=\frac{1}{2(L-2)}\sum_{j\neq a,b}\sum_{k_j=0,1}\{P[\sum_{i}\frac{(-1)^{k_i}\widetilde{c}_{ian}}{2^n}]+P[\sum_{i}\frac{(-1)^{k_i}\widetilde{c}_{ibn}}{2^n}]\}\\
&=P[\frac{(-1)^{k_a}\widetilde{c}_{aan}}{2^n}+\frac{(-1)^{k_b}\widetilde{c}_{ban}}{2^n}]+P[\frac{(-1)^{k_a}\widetilde{c}_{abn}}{2^n}+\frac{(-1)^{k_b}\widetilde{c}_{bbn}}{2^n}]+\sum_{i\neq a,b}\{\frac{c_{ian}^2}{4^n}P(\ket{e_{ian}})+\frac{c_{ibn}^2}{4^n}P(\ket{e_{ibn}}) \}.
\end{split}
\end{equation} 
If $k_a\oplus k_b=0$, $\rho_n^{a,b}$ will be
\begin{equation}
\rho_{0,n}^{a,b}=P[\frac{\widetilde{c}_{aan}}{2^n}+\frac{\widetilde{c}_{ban}}{2^n}]+P[\frac{\widetilde{c}_{abn}}{2^n}+\frac{\widetilde{c}_{bbn}}{2^n}]+\sum_{i\neq a,b}\{\frac{c_{ian}^2}{4^n}P(\ket{e_{ian}})+\frac{c_{ibn}^2}{4^n}P(\ket{e_{ibn}}) \}
\end{equation}
if $k_a\oplus k_b=1$, $\rho_n^{a,b}$ will be
\begin{equation}
\rho_{1,n}^{a,b}=P[\frac{\widetilde{c}_{aan}}{2^n}-\frac{\widetilde{c}_{ban}}{2^n}]+P[\frac{\widetilde{c}_{abn}}{2^n}-\frac{\widetilde{c}_{bbn}}{2^n}]+\sum_{i\neq a,b}\{\frac{c_{ian}^2}{4^n}P(\ket{e_{ian}})+\frac{c_{ibn}^2}{4^n}P(\ket{e_{ibn}}) \}.
\end{equation}
The mutual information of Alice and Eve if Bob announce $\{a,b\}$, $I_{AE}^{a,b}$, could be calculated by the Holevo bound
\begin{equation}
I_{AE}^{a,b}\le S[\sum_{n=odd}\frac{1}{2Q_{ab}}(\rho_{0,n}^{a,b}+\rho_{1,n}^{a,b})]-\frac{1}{2}S[\frac{1}{Q_{ab}}\sum_{n=odd}\rho_{0,n}^{a,b}]-\frac{1}{2}S[\frac{1}{Q_{ab}}\sum_{n=odd}\rho_{1,n}^{a,b}].
\end{equation}

Without compromising the security~\cite{yin2018improved}, we can assume
\begin{equation}\label{assum1}
\iprod{e_{ijn}}{e_{klm}}=\delta_{ik}\delta_{jl}\delta_{nm}.
\end{equation}
Then we could get the upper bound of $I_{AE}^{a,b}$
\begin{equation}
\begin{split}
Q_{ab}I_{AE}^{a,b}\le &\sum_{n=odd}\left(-\frac{c_{aan}^2}{4^n}log_2\frac{c_{aan}^2}{4^n}-\frac{c_{ban}^2}{4^n}log_2\frac{c_{ban}^2}{4^n}+\frac{c_{aan}^2+c_{ban}^2}{4^n}log_2\frac{c_{aan}^2+c_{ban}^2}{4^n}\right.\\
&\left. -\frac{c_{abn}^2}{4^n}log_2\frac{c_{abn}^2}{4^n}-\frac{c_{bbn}^2}{4^n}log_2\frac{c_{bbn}^2}{4^n}+\frac{c_{abn}^2+c_{bbn}^2}{4^n}log_2\frac{c_{abn}^2+c_{bbn}^2}{4^n}\right)\\
=&\sum_{n=odd}[\varphi(\frac{c_{aan}^2}{4^n},\frac{c_{ban}^2}{4^n})+\varphi(\frac{c_{bbn}^2}{4^n},\frac{c_{abn}^2}{4^n})],
\end{split}
\end{equation}
where
\begin{align}
Q_{ab}&=\sum_{n=odd}\sum_i(\frac{c_{ian}^2}{4^n}+\frac{c_{ibn}^2}{4^n})\\
\varphi(x,y)&=-xlog_2x-ylog_2(y)+(x+y)log_2(x+y)
\end{align}
If we denote $p_{ij}^2=\sum_{n=odd} \frac{c_{ijn}^2}{4^n}$, with the concavity of $\varphi(x,y)$ and Jensen’s inequality, we have
\begin{equation}\label{iae1}
Q_{ab}I_{AE}^{a,b}\le \varphi(p_{aa}^2,p_{ba}^2)+\varphi(p_{bb}^2,p_{ab}^2)
\end{equation}
where
\begin{equation}
Q_{ab}=\sum_i(p_{ia}^2+p_{ib}^2)
\end{equation}
The Eq.~\eqref{iae1} is the same as Eq. (7) of the security proof in the single-photon case of Ref.~\cite{yin2018improved}. The upper bound of $I_{AE}^{Odd}$ is
\begin{equation}
\begin{split}
I_{AE}^{Odd}&=\frac{\sum_{a<b}Q_{ab}I_{AE}^{a,b}}{\sum_{a<b}Q_{ab}}\\
&\le \frac{\sum_{a<b}\varphi(p_{aa}^2,p_{ba}^2)+\varphi(p_{bb}^2,p_{ab}^2)}{\sum_{a<b}Q_{ab}}\\
&\le \frac{\varphi(\sum_{a<b}p_{aa}^2+p_{bb}^2,\sum_{a<b}p_{ba}^2+p_{ab}^2)}{\sum_{a<b}Q_{ab}}\\
&=\frac{\varphi[(L-1)\sum_ip_{ii}^2,\sum_{i\neq j}p_{ij}^2]}{(L-1)(\sum_i p_{ii}^2+\sum_{i\neq j}p_{ij}^2)}\\
&=\frac{\varphi[(L-1)x_1,x_2]}{(L-1)(x_1+x_2)},
\end{split}
\end{equation}
where $x_1=\sum_i p_{ii}^2$ and $x_2=\sum_{i\neq j}p_{ij}^2$. By maximizing the value of $\frac{\varphi[(L-1)x_1,x_2]}{(L-1)(x_1+x_2)}$ for all $x_1>0$ and $x_2>0$, we could get the upper bound of $I_{AE}^{Odd}$. 
\subsubsection{The mutual information of Alice and Eve with even-count}
In this part, we want to prove that if the single-photon pulses train causes an even-count, Eve will have all information of Alice's raw key bits. Or in other word, the supremum of $I_{AE}^{Even}$ equals $1$. According to Eq.~\eqref{evestate}, if $n$ is an even number, the density matrix of Eve's ancilla bits will be
\begin{equation}
\rho_E=\sum_{n=even}\{P[\sum_{i}(-1)^{k_i}\frac{\widetilde{c}_{ian}}{2^n}+\sum_{i}(-1)^{k_i}\frac{\widetilde{c}_{ibn}}{2^n}]\}.
\end{equation}
We set $c_{ijn}=0$ if $n\neq 2$ or $i\neq j$, and $c_{ii2}=c$. Then $\rho_E$ will be the following simple form
\begin{equation}
\rho_E=P[(-1)^{k_i}\frac{\widetilde{c}_{aa2}}{4}+(-1)^{k_b}\frac{\widetilde{c}_{bb2}}{4}].
\end{equation} 
If $k_a\oplus k_b=0$,
\begin{equation}
\rho_E^0=P[\frac{\widetilde{c}_{aa2}}{4}+\frac{\widetilde{c}_{bb2}}{4}].
\end{equation} 
If $k_a\oplus k_b=1$,
\begin{equation}
\rho_E^1=P[\frac{\widetilde{c}_{aa2}}{4}-\frac{\widetilde{c}_{bb2}}{4}].
\end{equation} 
The mutual information of Alice and Eve if Bob announce $\{a,b\}$, $I_{AE}^{a,b}$, could be calculated by the Holevo bound
\begin{equation}
I_{AE}^{a,b}\le S[\frac{1}{2Q_{ab}}(\rho_E^0+\rho_E^1)]-\frac{1}{2}S[\frac{1}{Q_{ab}}\rho_E^0]-\frac{1}{2}S[\frac{1}{Q_{ab}}\rho_E^1]=1,
\end{equation}
where $Q_{ab}=\frac{c^2}{8}$. Here we use the orthogonality condition Eq.~\eqref{assum1}. Thus the upper bound of $I_{AE}^{even}$ is
\begin{equation}
I_{AE}^{Odd}=\frac{\sum_{a<b}Q_{ab}I_{AE}^{a,b}}{\sum_{a<b}Q_{ab}}\le 1.
\end{equation}
The only inequality we use here is Holevo bound, thus the supremum of $I_{AE}^{Even}$ equals $1$.
\subsubsection{The mutual information of Bob and Eve}
If the single photon pulses train causes a count in detector $D_1$, the density matrix of Eve is
\begin{equation}
\rho_E^{D_1}=\sum_{n=1}P[\sum_{i}(-1)^{k_i}\frac{\widetilde{c}_{ian}}{2^n}+\sum_{i}(-1)^{k_i}\frac{\widetilde{c}_{ibn}}{2^n}].
\end{equation} 
If the single photon pulses train causes a count in detector $D_2$, the density matrix of Eve is
\begin{equation}
\rho_E^{D_2}=\sum_{n=1}P[\sum_{i}(-1)^{k_i}\frac{\widetilde{c}_{ian}}{2^n}+\sum_{i}(-1)^{k_i}(-1)^n\frac{\widetilde{c}_{ibn}}{2^n}].
\end{equation}  
It is easy to see that $\rho_E^{D_1}$ and $\rho_E^{D_2}$ are the same if $n$ is even, which means Eve can not distinguish whether an even-count is caused by $D_1$ or $D_2$ at all, thus $I_{BE}^{Even}=0$. Different from the situation of $I_{AE}^{Odd}$, this result need not the average of random phase and orthogonality assumption Eq.~\eqref{assum1}.
Similar to the case of mutual information of Alice and Eve with even-counts, we want to prove that if the single-photon pulses train causes an odd-count, Eve will have all information of Bob's raw key bits. If $n$ is odd, $\rho_E^{D_1}$ and $\rho_E^{D_2}$ will be 
\begin{align}
\rho_E^{D_1}&=\sum_{n=odd}P[\sum_{i}(-1)^{k_i}\frac{\widetilde{c}_{ian}}{2^n}+\sum_{i}(-1)^{k_i}\frac{\widetilde{c}_{ibn}}{2^n}],\\
\rho_E^{D_2}&=\sum_{n=odd}P[\sum_{i}(-1)^{k_i}\frac{\widetilde{c}_{ian}}{2^n}-\sum_{i}(-1)^{k_i}\frac{\widetilde{c}_{ibn}}{2^n}].
\end{align}
If we set $c_{ijn}=0$ if $n\neq 1$ or $i\neq j$, and $c_{ii1}=c$, $\rho_E^{D_1}$ and $\rho_E^{D_2}$ will be 
\begin{equation}
\rho_E^{D_1}= P[\frac{\widetilde{c}_{aa1}}{2}+(-1)^{k_a+k_b}\frac{\widetilde{c}_{bb1}}{2}], \rho_E^{D_2}=  P[\frac{\widetilde{c}_{aa1}}{2}-(-1)^{k_a+k_b}\frac{\widetilde{c}_{bb1}}{2}].
\end{equation}
Since the states of $\rho_E^{D_1}$ and $\rho_E^{D_2}$ with assumption Eq.~\eqref{assum1} are two orthometric pure states, $I_{BE}^{a,b}\le 1$. And thus $I_{BE}^{Odd}\le 1$. 
\section{The upper bound of leakage information in general case}
In this part we want to evaluate the upper bound of leakage information in general case. We have shown that if the single-photon pulses train causes an even-count, the mutual information of Alice and Eve is up to $1$, and if the single-photon pulses train causes an odd-count, the mutual information of Bob and Eve is up to $1$. Honestly speaking, this is a trivial conclusion, thus we will focus on the upper bounds of mutual information of Alice and Eve if the pulses train causes an odd-count and the mutual information of Bob and Eve if the pulses train causes an even-count. Same as the method in Ref.~\cite{yin2018improved}, we will prove this in two cases, $N$ is an odd number and $N$ is an even number, where $N$ denotes Alice sends out a $N$-photons $L$-pulses train. We assume $N\le L/2$ in this part. 
\subsection{The upper bound of leakage information if N is an odd number}
If Alice prepares a $N$-photon $L$-pulses train where $N$ is an odd number, the state will be
\begin{equation}\label{eq10}
\ket{\Psi}_A=\sum_{i_1}(-1)^{k_{i_1}}\ket{i_1}+\sum_{i_1<i_2<i_3}(-1)^{k_{i_1}+k_{i_2}+k_{i_3}}\ket{i_1i_2i_3}+\cdots+\sum_{i_1<i_2<i_3<\cdots<i_N}(-1)^{k_{i_1}+k_{i_2}+k_{i_3}+\cdots+k_{i_N}}\ket{i_1i_2i_3\cdots i_N},
\end{equation}
where $\ket{i_1i_2i_3\cdots i_k}$ represents a sum of all states that there are odd number photons in the $i_1,i_2,i_3,\dots,i_k$-th pulses and there are even number photons in the other pulses of the $L$-pulses train, for $k=1,3,5,\dots,N$. 
Eve's optimal collective attack can be given by the following equation for the state shown in Eq.~\eqref{eq10}:
\begin{equation}\label{eq34}
U_{eve}\ket{i_1i_2i_3\cdots i_k}\ket{e_{ancilla}}=\sum_{n=0}^\infty\sum_{t=1}^L c_{i_1i_2\cdots i_ktn}\ket{n_t}\ket{e_{i_1i_2\cdots i_ktn}},
\end{equation}
and we denote
\begin{equation}\label{eq35}
\widetilde{c}_{i_1i_2\cdots i_ktn}\equiv c_{i_1i_2\cdots i_ktn}\ket{e_{i_1i_2\cdots i_ktn}},
\end{equation}
for $k=1,3,5,\dots,N$.
Then Alice and Eve will share the following entanglement state
\begin{equation}\label{eq11}
\begin{split}
\ket{\Psi}_{AE}=&\sum_{n=0}^\infty\sum_{t=1}^L\left[\sum_{i_1}(-1)^{k_{i_1}}\widetilde{c}_{i_1tn}+\sum_{i_1<i_2<i_3}(-1)^{k_{i_1}+k_{i_2}+k_{i_3}}\widetilde{c}_{i_1i_2i_3tn}+\cdots\right.\\
&+\sum_{i_1<i_2<i_3<\cdots<i_N}(-1)^{k_{i_1}+k_{i_2}+k_{i_3}+\cdots+k_{i_N}}\widetilde{c}_{i_1i_2i_3\cdots i_Ntn}\left. \right]\ket{n_t}
\end{split}
\end{equation}
Further, we denote
\begin{equation}
\begin{split}
\widetilde{c}_{tn}=\sum_{i_1}(-1)^{k_{i_1}}\widetilde{c}_{i_1tn}+\sum_{i_1<i_2<i_3}(-1)^{k_{i_1}+k_{i_2}+k_{i_3}}\widetilde{c}_{i_1i_2i_3tn}+\cdots+\sum_{i_1<i_2<i_3<\cdots<i_N}(-1)^{k_{i_1}+k_{i_2}+k_{i_3}+\cdots+k_{i_N}}\widetilde{c}_{i_1i_2i_3\cdots i_Ntn},
\end{split}
\end{equation}
Through Bob's detection set-up, the state of Eq.~\eqref{eq11} is involved into 
\begin{align}
\ket{\Psi}_{AE}=&\sum_{n=0}^\infty\sum_{t=1}^L\widetilde{c}_{tn}\ket{n_t}\\
&\to \sum_{tn}\widetilde{c}_{tn}\frac{\hat{d}_t^{\dagger,n}}{\sqrt{n!}}\ket{0}\\
&\to \sum_{tn}\widetilde{c}_{tn}\frac{1}{2^n\sqrt{n!}}(\hat{d}_{t+r}^{\dagger}+\hat{e}_{t+r}^{\dagger}+\hat{d}_{t}^{\dagger}-\hat{e}_{t}^{\dagger})^n\ket{0}\equiv \ket{\Psi}_{AE}^{detect}.
\end{align} 
If only one of Bob's detectors responses in location $k$ and we denote $a=k-r,b=k$, the density matrix of Eve's ancilla bits is
\begin{equation}\label{eq12}
\rho_E=\frac{1}{2}\sum_{n=1}\left\{P(\widetilde{c}_{an}+\widetilde{c}_{bn})+P[\widetilde{c}_{an}+(-1)^n\widetilde{c}_{bn}]\right\}
\end{equation} 
The first part of Eq.~\eqref{eq12} is caused by detector $D_1$, and the second part is caused by detector $D_2$. It is easy to see that those two parts are the same if $n$ is an even and thus $I_{BE}^{Even}(N)$=0. If $n$ is odd, we have the following consideration to evaluate the upper bound of mutual information of Alice and Eve.
\begin{equation}\label{eq442}
\rho_E=\sum_{n=odd}P(\widetilde{c}_{an})+P(\widetilde{c}_{bn}).
\end{equation}
We first transform $P(\widetilde{c}_{an})$ and $P(\widetilde{c}_{bn})$ into
\begin{equation}
\begin{split}
P(\widetilde{c}_{an})=&P\{\sum_{i_1}(-1)^{k_{i_1}}\widetilde{c}_{i_1an}+\sum_{i_1<i_2<i_3}(-1)^{k_{i_1}+k_{i_2}+k_{i_3}}\widetilde{c}_{i_1i_2i_3an}+\cdots\\
&+\sum_{i_1<i_2<i_3<\cdots<i_N}(-1)^{k_{i_1}+k_{i_2}+k_{i_3}+\cdots +k_{i_N}}\widetilde{c}_{i_1i_2i_3\cdots i_Nan}\}\\
=&P\{(-1)^{k_a}\widetilde{c}_{aan}+(-1)^{k_b}\widetilde{c}_{ban}\\ 
&+\sum_{i_1\neq a,b}(-1)^{k_i}[\widetilde{c}_{i_1an}+(-1)^{k_a+k_b}\widetilde{c}_{i_1aban}]\\
&+\sum_{i_1<i_2,i_1,i_2\neq a,b}(-1)^{k_{i_1}+k_{i_2}}[(-1)^{k_a}\widetilde{c}_{i_1i_2aan}+(-1)^{k_b}\widetilde{c}_{i_1i_2ban}]\\
&+\sum_{\substack{i_1<i_2<i_3\\ i_1,i_2,i_3\neq a,b}}(-1)^{k_{i_1}+k_{i_2}+k_{i_3}}[(-1)^{k_a+k_b}\widetilde{c}_{i_1i_2i_3aban}+\widetilde{c}_{i_1i_2i_3an}]\\
&+\sum_{\substack{i_1<i_2<i_3<i_4\\i_1,i_2,i_3,i_4\neq a,b}}(-1)^{k_{i_1}+k_{i_2}+k_{i_3}+k_{i_4}}[(-1)^{k_a}\widetilde{c}_{i_1i_2i_3i_4aan}+(-1)^{k_b}\widetilde{c}_{i_1i_2i_3i_4ban}]\\
&+\dots\dots\\
&+\sum_{i_1<i_2<i_3<\cdots<i_{N-1}}(-1)^{k_{i_1}+k_{i_2}+k_{i_3}+\cdots+k_{i_{N-1}}}[(-1)^{k_a}\widetilde{c}_{i_1i_2i_3\cdots i_{N-1}aan}+(-1)^{k_b}\widetilde{c}_{i_1i_2i_3\cdots i_{N-1}ban}]\\
&\sum_{\substack{i_1<i_2<i_3<\cdots<i_N\\i_1,i_2,\cdots,i_N\neq a,b}}(-1)^{k_{i_1}+k_{i_2}+k_{i_3}+\cdots +k_{i_N}}\widetilde{c}_{i_1i_2i_3\cdots i_Nan}
\end{split}
\end{equation}
The value of $k_i,(i\neq a,b)$ is randomly $0$ or $1$, thus we have
\begin{equation}\label{eq44}
\begin{split}
P(\widetilde{c}_{an})&=P[(-1)^{k_a}\widetilde{c}_{aan}+(-1)^{k_b}\widetilde{c}_{ban}]+\sum_{i_1\neq a,b}P[\widetilde{c}_{i_1an}+(-1)^{k_a+k_b}\widetilde{c}_{i_1aban}]\\
&+\sum_{i_1<i_2,i_1,i_2\neq a,b}P[(-1)^{k_a}\widetilde{c}_{i_1i_2aan}+(-1)^{k_b}\widetilde{c}_{i_1i_2ban}]+\sum_{\substack{i_1<i_2<i_3\\ i_1,i_2,i_3\neq a,b}}P[(-1)^{k_a+k_b}\widetilde{c}_{i_1i_2i_3aban}+\widetilde{c}_{i_1i_2i_3an}]\\
&+\sum_{\substack{i_1<i_2<i_3<i_4\\i_1,i_2,i_3,i_4\neq a,b}}P[(-1)^{k_a}\widetilde{c}_{i_1i_2i_3i_4aan}+(-1)^{k_b}\widetilde{c}_{i_1i_2i_3i_4ban}]+\dots\dots\\
&+\sum_{\substack{i_1<i_2<i_3<\cdots<i_{N-1}\\i_1,i_2,\cdots,i_{N-1}\neq a,b}}P[(-1)^{k_a}\widetilde{c}_{i_1i_2i_3\cdots i_{N-1}aan}+(-1)^{k_b}\widetilde{c}_{i_1i_2i_3\cdots i_{N-1}ban}]\\
&+\sum_{\substack{i_1<i_2<i_3<\cdots<i_N\\i_1,i_2,\cdots,i_N\neq a,b}}P(\widetilde{c}_{i_1i_2i_3\cdots i_Nan})
\end{split}
\end{equation}
Similarly, $P(\widetilde{c}_{bn})$ could be transformed into the following form
\begin{equation}\label{eq45}
\begin{split}
P(\widetilde{c}_{bn})&=P[(-1)^{k_a}\widetilde{c}_{abn}+(-1)^{k_b}\widetilde{c}_{bbn}]+\sum_{i_1\neq a,b}P[\widetilde{c}_{i_1bn}+(-1)^{k_a+k_b}\widetilde{c}_{i_1abbn}]\\
&+\sum_{i_1<i_2,i_1,i_2\neq a,b}P[(-1)^{k_a}\widetilde{c}_{i_1i_2abn}+(-1)^{k_b}\widetilde{c}_{i_1i_2bbn}]+\sum_{\substack{i_1<i_2<i_3\\ i_1,i_2,i_3\neq a,b}}P[(-1)^{k_a+k_b}\widetilde{c}_{i_1i_2i_3abbn}+\widetilde{c}_{i_1i_2i_3bn}]\\
&+\sum_{\substack{i_1<i_2<i_3<i_4\\i_1,i_2,i_3,i_4\neq a,b}}P[(-1)^{k_a}\widetilde{c}_{i_1i_2i_3i_4abn}+(-1)^{k_b}\widetilde{c}_{i_1i_2i_3i_4bbn}]+\dots\dots\\
&+\sum_{\substack{i_1<i_2<i_3<\cdots<i_{N-1}\\i_1,i_2,\cdots,i_{N-1}\neq a,b}}P[(-1)^{k_a}\widetilde{c}_{i_1i_2i_3\cdots i_{N-1}abn}+(-1)^{k_b}\widetilde{c}_{i_1i_2i_3\cdots i_{N-1}bbn}]\\
&+\sum_{\substack{i_1<i_2<i_3<\cdots<i_N\\i_1,i_2,\cdots,i_N\neq a,b}}P(\widetilde{c}_{i_1i_2i_3\cdots i_Nbn})
\end{split}
\end{equation}
We denote $\rho_E$ as $\rho_E^0$ if $k_a\oplus k_b=0$ and $\rho_E$ as $\rho_E^1$ if $k_a\oplus k_b=1$, thus the mutual information of Alice and Eve if Bob announces $\{a,b\}$, $I_{AE}^{a,b}$ is
\begin{equation}
I_{A,E}^{a,b}\le S[\frac{1}{2Q_{ab}}(\rho_E^0+\rho_E^1)]-\frac{1}{2}S[\frac{\rho_E^0}{Q_{ab}}]-\frac{1}{2}S[\frac{\rho_E^1}{Q_{ab}}],
\end{equation}
where
\begin{equation}
Q_{ab}=\sum_{n=odd}[\sum_{i_1}(|\widetilde{c}_{i_1an}|^2+|\widetilde{c}_{i_1bn}|^2)+\sum_{i_1<i_2<i_3}(|\widetilde{c}_{i_1i_2i_3an}|^2+|\widetilde{c}_{i_1i_2i_3bn}|^2)+\cdots+\sum_{i_1<i_2<i_3<\cdots<i_N}(|\widetilde{c}_{i_1i_2i_3\cdots i_Nan}|^2+|\widetilde{c}_{i_1i_2i_3\cdots i_Nbn}|^2)].
\end{equation}
We denote $p_{i_1i_2i_3\cdots i_kt}^2=\sum_{n=odd}|\widetilde{c}_{i_1i_2i_3\cdots i_ktn}|^2$ where $k=1,2,3,\dots,N$. And the upper bound of $I_{AE}$ is 
\begin{equation}
\begin{split}
I_{AE}=&\frac{\sum_{a<b}Q_{ab}I_{AE}^{a,b}}{\sum_{a<b}Q_{ab}}\\
\le&\frac{1}{\sum_{a<b}Q_{ab}} \sum_{a<b}\sum_{n=odd}[\varphi(|\widetilde{c}_{aan}|^2+|\widetilde{c}_{bbn}|^2,|\widetilde{c}_{abn}|^2+|\widetilde{c}_{ban}|^2)\\
&+\sum_{i_1\neq a,b}\varphi(|\widetilde{c}_{i_1an}|^2+|\widetilde{c}_{i_1bn}|^2,|\widetilde{c}_{i_1aban}|^2+|\widetilde{c}_{i_1abbn}|^2)\\
&+\sum_{i_1<i_2,i_1,i_2\neq a,b}\varphi(|\widetilde{c}_{i_1i_2aan}|^2+|\widetilde{c}_{i_1i_2bbn}|^2,|\widetilde{c}_{i_1i_2abn}|^2+|\widetilde{c}_{i_1i_2ban}|^2)\\
&+\sum_{\substack{i_1<i_2<i_3\\i_1,i_2i_3\neq a,b}}\varphi(|\widetilde{c}_{i_1i_2i_3an}|^2+|\widetilde{c}_{i_1i_2i_3bn}|^2,|\widetilde{c}_{i_1i_2i_3aban}|^2+|\widetilde{c}_{i_1i_2i_3abbn}|^2)+\dots\dots\\
&+\sum_{\substack{i_1<i_2<i_3<\cdots<i_{N-1}\\i_1,i_2,\cdots,i_{N-1}\neq a,b}}\varphi(|\widetilde{c}_{i_1i_2i_3\cdots i_{N-1}aan}|^2+|\widetilde{c}_{i_1i_2i_3\cdots i_{N-1}bbn}|^2,\widetilde{c}_{i_1i_2i_3\cdots i_{N-1}abn}|^2+|\widetilde{c}_{i_1i_2i_3\cdots i_{N-1}ban}|^2)]\\
\le&\frac{1}{\sum_{a<b}Q_{ab}} \sum_{a<b}[\varphi(p_{aa}^2+p_{bb}^2,p_{ab}^2+p_{ba}^2)+\sum_{i_1\neq a,b}\varphi(p_{i_1a}^2+p_{i_1b}^2,p_{i_1aba}^2+p_{i_1abb}^2)\\
&+\sum_{i_1<i_2,i_1,i_2\neq a,b}\varphi(p_{i_1i_2aa}^2+p_{i_1i_2bb}^2,p_{i_1i_2ab}^2+p_{i_1i_2ba}^2)+\sum_{\substack{i_1<i_2<i_3\\i_1,i_2i_3\neq a,b}}\varphi(p_{i_1i_2i_3a}^2+p_{i_1i_2i_3b}^2,p_{i_1i_2i_3aba}^2+p_{i_1i_2i_3abb}^2)\\
&+\dots\dots+\sum_{\substack{i_1<i_2<i_3<\cdots<i_{N-1}\\i_1,i_2,\cdots,i_{N-1}\neq a,b}}\varphi(p_{i_1i_2i_3\cdots i_{N-1}aa}^2+p_{i_1i_2i_3\cdots i_{N-1}bb}^2,p_{i_1i_2i_3\cdots i_{N-1}ab}^2+p_{i_1i_2i_3\cdots i_{N-1}ba}^2)].
\end{split}
\end{equation}
As shown in Eq. (44) of supplementary note of Ref.~\cite{yin2018improved}, we have the following mathematical identities
\begin{align*}
&\sum_{a<b}p_{aa}^2+p_{bb}^2=(L-1)\sum_{i_1}p_{i_1i_1}^2\\
&\sum_{a<b}p_{ab}^2+p_{ba}^2=\sum_{i_1}\sum_{i_2\neq i_1}p_{i_1i_2}^2\\
&\sum_{a<b}\sum_{i_1\neq a,b}p_{i_1a}^2+p_{i_1b}^2=(L-2)\sum_{i_1}\sum_{i_2\neq i_1}p_{i_1i_2}^2\\
&\sum_{a<b}\sum_{\substack{i_1<i_2<\cdots<i_k\\i_1,i_2,\dots,i_k\neq a,b}}p_{i_1i_2\cdots i_ka}^2+p_{i_1i_2\cdots i_kb}^2=(L-k-1)\sum_{i_1<i_2<\cdots<i_k}\sum_{t\neq i_1,i_2,\dots,i_k}p_{i_1i_2\cdots i_kt}^2\\
&\sum_{a<b}\sum_{\substack{i_1<i_2<\cdots<i_k\\i_1,i_2,\dots,i_k\neq a,b}}p_{i_1i_2\cdots i_kaba}^2+p_{i_1i_2\cdots i_kabb}^2=(k+1)\sum_{i_1<i_2<\cdots<i_{k+2}}\sum_{t=i_1}^{i_{k+2}}p_{i_1i_2\cdots i_{k+2}t}^2\\
&\sum_{a<b}\sum_{\substack{i_1<i_2<\cdots<i_{k+1}\\i_1,i_2,\dots,i_{k+1}\neq a,b}}p_{i_1i_2\cdots i_{k+1}aa}^2+p_{i_1i_2\cdots i_{k+1}bb}^2=(L-k-2)\sum_{i_1<i_2<\cdots<i_{k+2}}\sum_{t=i_1}^{i_{k+2}}p_{i_1i_2\cdots i_{k+2}t}^2\\
&\sum_{a<b}\sum_{\substack{i_1<i_2<\cdots<i_{k+1}\\i_1,i_2,\dots,i_{k+1}\neq a,b}}p_{i_1i_2\cdots i_{k+1}ab}^2+p_{i_1i_2\cdots i_{k+1}ba}^2=(k+2)\sum_{i_1<i_2<\cdots<i_{k+2}}\sum_{t\neq i_1\cdots,i_{k+2}}p_{i_1i_2\cdots i_{k+2}t}^2,\\
\end{align*}
where $k=1,3,5,\dots,N$.
And we define
\begin{align*}
&x_1\equiv\sum_{i_1}p_{i_1i_1}^2\\
&x_2\equiv\sum_{i_1\neq i_2}p_{i_1i_2}^2\\
&x_k\equiv\sum_{i_1<i_2<\cdots<i_k}\sum_{t=i_1}^{i_k}p_{i_1i_2\cdots i_kt}^2\\
&x_{k+1}\equiv\sum_{i_1<i_2<\cdots<i_k}\sum_{t\neq i_1,\dots,i_k}p_{i_1i_2\cdots i_kt}^2
\end{align*}
It is easy to check $\sum_{a<b}Q_{ab}=(L-1)(x_1+x_2+\cdots+x_{N+1})$. With the concavity of $\varphi(x,y)$ and Jensen’s inequality, we have 
\begin{equation}\label{eq49}
I_{AE}\le\frac{\sum_{k=1}^N\varphi[(L-k)x_k,kx_{k+1}]}{(L-1)\sum_{k=1}^{N+1}x_k},
\end{equation}
By searching the maximum value of Eq.~\eqref{eq49} in non-negative spaces,we could get the upper bound of $I_{AE}$.
\subsection{The upper bound of leakage information if N is an even number}
If Alice prepares a $N$-photon $L$-pulses train where $N$ is an even number, the state will be
\begin{equation}
\begin{split}
\ket{\Psi}_A=&\ket{i_1}+\sum_{i_1<i_2}(-1)^{k_{i_1}+k_{i_2}}\ket{i_1i_2}+\sum_{i_1<i_2<i_3<i_4}(-1)^{k_{i_1}+k_{i_2}+k_{i_3}+k_{i_4}}\ket{i_1i_2i_3i_4}+\cdots\\
&+\sum_{i_1<i_2<i_3<\cdots<i_N}(-1)^{k_{i_1}+k_{i_2}+k_{i_3}+\cdots+k_{i_N}}\ket{i_1i_2i_3\cdots i_N},
\end{split}
\end{equation}
where $\ket{i_1i_2i_3\cdots i_k}$ represents a sum of all states that there are odd number photons in the $i_1,i_2,i_3,\dots,i_k$-th pulses and there are even number photons in the other pulses of the $L$-pulses train, for $k=2,4,6,\dots,N$. And $\ket{i_1}$ represents a sum of all states that there are only even number photons in any pulse. Similarly to Eqs.~\eqref{eq34} and~\eqref{eq35}, Eve's optimal collective attack is 
\begin{equation}
U_{eve}\ket{i_1i_2\cdots i_k}\ket{e_{ancilla}}=\sum_{n=0}^{\infty}\sum_{t=1}^L\widetilde{c}_{i_1i_2\cdots i_k tn}\ket{n_t},
\end{equation}
for $k=1,2,4,6,\dots,N$. And similarly to Eqs.~\eqref{eq11}-\eqref{eq12}, if only one of Bob's detectors responses in location $k$ and we denote $a=k-r,b=k$, the density matrix of Eve's ancilla bits is
\begin{equation}\label{eq53}
\rho_E=\frac{1}{2}\sum_{n=1}\left\{P(\widetilde{c}_{an}+\widetilde{c}_{bn})+P[\widetilde{c}_{an}+(-1)^n\widetilde{c}_{bn}]\right\}
\end{equation} 
The first part of Eq.~\eqref{eq53} is caused by detector $D_1$, and the second part is caused by detector $D_2$. It is easy to see that those two parts are the same if $n$ is even and thus $I_{BE}^{Even}(N)$=0. If $n$ is odd, we have the following consideration to evaluate the upper bound of mutual information of Alice and Eve.
\begin{equation}\label{eq42}
\rho_E=\sum_{n=odd}P(\widetilde{c}_{an})+P(\widetilde{c}_{bn}),
\end{equation}
where 
\begin{equation}
\begin{split}
P(\widetilde{c}_{an})=&P\{\widetilde{c}_{i_1an}+\sum_{i_1<i_2}(-1)^{k_{i_1}+k_{i_2}}\widetilde{c}_{i_1i_2an}\ket{i_1i_2}+\sum_{i_1<i_2<i_3<i_4}(-1)^{k_{i_1}+k_{i_2}+k_{i_3}+k_{i_4}}\widetilde{c}_{i_1i_2i_3i_4an}\ket{i_1i_2i_3i_4}+\cdots\\
&+\sum_{i_1<i_2<i_3<\cdots<i_N}(-1)^{k_{i_1}+k_{i_2}+k_{i_3}+\cdots+k_{i_N}}\widetilde{c}_{i_1i_2i_3\cdots i_Nan}\ket{i_1i_2i_3\cdots i_N}\}\\
=&P\{\widetilde{c}_{i_1an}+(-1)^{k_{a}+k_{b}}\widetilde{c}_{aban}\\
&+\sum_{i_1\neq a,b}(-1)^{k_i}[(-1)^{k_a}\widetilde{c}_{i_1aan}+(-1)^{k_b}\widetilde{c}_{i_1ban}]\\
&+\sum_{i_1<i_2,i_1,i_2\neq a,b}(-1)^{k_{i_1}+k_{i_2}}(\widetilde{c}_{i_1i_2an}+(-1)^{k_a+k_b}\widetilde{c}_{i_1i_2aban})\\
&+\sum_{\substack{i_1<i_2<i_3\\i_1,i_2,i_3\neq a,b}}(-1)^{k_{i_1}+k_{i_2}+k_{i_3}}[(-1)^{k_a}\widetilde{c}_{i_1i_2i_3aan}+(-1)^{k_b}\widetilde{c}_{i_1i_2i_3ban}]+\cdots\\
&+\sum_{\substack{i_1<i_2<i_3<\cdots<i_{N-1}\\i_1,i_2,\dots i_{N-1}\neq a,b}}(-1)^{k_{i_1}+k_{i_2}+\cdots+k_{i_{N-1}}}[(-1)^{k_a}\widetilde{c}_{i_1i_2\cdots i_{N-1}aan}+(-1)^{k_b}\widetilde{c}_{i_1i_2\cdots i_{N-1}ban}]\\
&+\sum_{\substack{i_1<i_2<i_3<\cdots<i_{N}\\i_1,i_2,\dots i_{N}\neq a,b}}(-1)^{k_{i_1}+k_{i_2}+\cdots+k_{i_{N}}}\widetilde{c}_{i_1i_2\cdots i_{N}an}\}
\end{split}
\end{equation}
The value of $k_i,(i\neq a,b)$ is randomly $0$ or $1$, thus we have
\begin{equation}
\begin{split}
P(\widetilde{c}_{an})=&P[\widetilde{c}_{i_1an}+(-1)^{k_{a}+k_{b}}\widetilde{c}_{aban}]+\sum_{i_1\neq a,b}P[(-1)^{k_a}\widetilde{c}_{i_1aan}+(-1)^{k_b}\widetilde{c}_{i_1ban}]\\
&+\sum_{i_1<i_2,i_1,i_2\neq a,b}P[\widetilde{c}_{i_1i_2an}+(-1)^{k_a+k_b}\widetilde{c}_{i_1i_2aban}]+\sum_{\substack{i_1<i_2<i_3\\i_1,i_2,i_3\neq a,b}}P[(-1)^{k_a}\widetilde{c}_{i_1i_2i_3aan}+(-1)^{k_b}\widetilde{c}_{i_1i_2i_3ban}]\\
&+\cdots+\sum_{\substack{i_1<i_2<i_3<\cdots<i_{N-1}\\i_1,i_2,\dots i_{N-1}\neq a,b}}P[(-1)^{k_a}\widetilde{c}_{i_1i_2\cdots i_{N-1}aan}+(-1)^{k_b}\widetilde{c}_{i_1i_2\cdots i_{N-1}ban}]+\sum_{\substack{i_1<i_2<i_3<\cdots<i_{N}\\i_1,i_2,\dots i_{N}\neq a,b}}P(\widetilde{c}_{i_1i_2\cdots i_{N}an})
\end{split}
\end{equation}
Similarly,
\begin{equation}
\begin{split}
P(\widetilde{c}_{bn})=&P[\widetilde{c}_{i_1bn}+(-1)^{k_{a}+k_{b}}\widetilde{c}_{abbn}]+\sum_{i_1\neq a,b}P[(-1)^{k_a}\widetilde{c}_{i_1abn}+(-1)^{k_b}\widetilde{c}_{i_1bbn}]\\
&+\sum_{i_1<i_2,i_1,i_2\neq a,b}P[\widetilde{c}_{i_1i_2bn}+(-1)^{k_a+k_b}\widetilde{c}_{i_1i_2abbn}]+\sum_{\substack{i_1<i_2<i_3\\i_1,i_2,i_3\neq a,b}}P[(-1)^{k_a}\widetilde{c}_{i_1i_2i_3abn}+(-1)^{k_b}\widetilde{c}_{i_1i_2i_3bbn}]\\
&+\cdots+\sum_{\substack{i_1<i_2<i_3<\cdots<i_{N-1}\\i_1,i_2,\dots i_{N-1}\neq a,b}}P[(-1)^{k_a}\widetilde{c}_{i_1i_2\cdots i_{N-1}abn}+(-1)^{k_b}\widetilde{c}_{i_1i_2\cdots i_{N-1}bbn}]+\sum_{\substack{i_1<i_2<i_3<\cdots<i_{N}\\i_1,i_2,\dots i_{N}\neq a,b}}P(\widetilde{c}_{i_1i_2\cdots i_{N}bn})
\end{split}
\end{equation}
We denote $p_{i_1i_2i_3\cdots i_kt}^2=\sum_{n=odd}|\widetilde{c}_{i_1i_2i_3\cdots i_ktn}|^2$ where $k=1,2,4,6,\dots,N$. And the upper bound of $I_{AE}$ is 
\begin{equation}
\begin{split}
I_{AE}=&\frac{\sum_{a<b}Q_{ab}I_{AE}^{a,b}}{\sum_{a<b}Q_{ab}}\\
\le&\frac{1}{\sum_{a<b}Q_{ab}} \sum_{a<b}\sum_{n=odd}[\varphi(|\widetilde{c}_{i_1an}|^2+|\widetilde{c}_{i_1bn}|^2,|\widetilde{c}_{aban}|^2+|\widetilde{c}_{abbn}|^2)\\
&+\sum_{i_1\neq a,b}\varphi(|\widetilde{c}_{i_1aan}|^2+|\widetilde{c}_{i_1bbn}|^2,|\widetilde{c}_{i_1ban}|^2+|\widetilde{c}_{i_1bbn}|^2)\\
&+\sum_{i_1<i_2,i_1,i_2\neq a,b}\varphi(|\widetilde{c}_{i_1i_2an}|^2+|\widetilde{c}_{i_1i_2bn}|^2,|\widetilde{c}_{i_1i_2aban}|^2+|\widetilde{c}_{i_1i_2abbn}|^2)\\
&+\sum_{\substack{i_1<i_2<i_3\\i_1,i_2i_3\neq a,b}}\varphi(|\widetilde{c}_{i_1i_2i_3aan}|^2+|\widetilde{c}_{i_1i_2i_3bbn}|^2,|\widetilde{c}_{i_1i_2i_3abn}|^2+|\widetilde{c}_{i_1i_2i_3ban}|^2)+\dots\dots\\
&+\sum_{\substack{i_1<i_2<i_3<\cdots<i_{N-1}\\i_1,i_2,\cdots,i_{N-1}\neq a,b}}\varphi(|\widetilde{c}_{i_1i_2i_3\cdots i_{N-1}aan}|^2+|\widetilde{c}_{i_1i_2i_3\cdots i_{N-1}bbn}|^2,\widetilde{c}_{i_1i_2i_3\cdots i_{N-1}abn}|^2+|\widetilde{c}_{i_1i_2i_3\cdots i_{N-1}ban}|^2)]\\
\le&\frac{1}{\sum_{a<b}Q_{ab}} \sum_{a<b}[\varphi(p_{i_1a}^2+p_{i_1b}^2,p_{aba}^2+p_{abb}^2)+\sum_{i_1\neq a,b}\varphi(p_{i_1aa}^2+p_{i_1bb}^2,p_{i_1ba}^2+p_{i_1ab}^2)\\
&+\sum_{i_1<i_2,i_1,i_2\neq a,b}\varphi(p_{i_1i_2a}^2+p_{i_1i_2b}^2,p_{i_1i_2abb}^2+p_{i_1i_2aba}^2)+\sum_{\substack{i_1<i_2<i_3\\i_1,i_2i_3\neq a,b}}\varphi(p_{i_1i_2i_3aa}^2+p_{i_1i_2i_3bb}^2,p_{i_1i_2i_3ab}^2+p_{i_1i_2i_3ba}^2)\\
&+\dots\dots+\sum_{\substack{i_1<i_2<i_3<\cdots<i_{N-1}\\i_1,i_2,\cdots,i_{N-1}\neq a,b}}\varphi(p_{i_1i_2i_3\cdots i_{N-1}aa}^2+p_{i_1i_2i_3\cdots i_{N-1}bb}^2,p_{i_1i_2i_3\cdots i_{N-1}ab}^2+p_{i_1i_2i_3\cdots i_{N-1}ba}^2)],
\end{split}
\end{equation}
where
\begin{equation}
\begin{split}
Q_{ab}=&\sum_{n=odd}[(|\widetilde{c}_{i_1an}|^2+|\widetilde{c}_{i_1bn}|^2)+\sum_{i_1<i_2}(|\widetilde{c}_{i_1i_2an}|^2+|\widetilde{c}_{i_1i_2bn}|^2)+\sum_{i_1<i_2<i_3<i_4}(|\widetilde{c}_{i_1i_2i_3i_4an}|^2+|\widetilde{c}_{i_1i_2i_3i_4bn}|^2)\\
&+\cdots+\sum_{i_1<i_2<i_3<\cdots<i_N}(|\widetilde{c}_{i_1i_2i_3\cdots i_Nan}|^2+|\widetilde{c}_{i_1i_2i_3\cdots i_Nbn}|^2)]\\
=&p_{i_1a}^2+p_{i_1b}^2+\sum_{i_1<i_2}(p_{i_1i_2a}^2+p_{i_1i_2b}^2)+\sum_{i_1<i_2<i_3<i_4}(p_{i_1i_2i_3i_4a}^2+p_{i_1i_2i_3i_4b}^2)\\
&+\cdots+\sum_{i_1<i_2<i_3<\cdots<i_N}(p_{i_1i_2\dots i_Na}^2+p_{i_1i_2\dots i_Nb}^2)
\end{split}
\end{equation}
As shown in Eq. (54) of supplementary note of Ref.~\cite{yin2018improved}, we have the following mathematical identities
\begin{align*}
&\sum_{a<b}p_{i_1a}^2+p_{i_1b}^2=(L-1)\sum_{i_2}p_{i_1i_2}^2\\
&\sum_{a<b}p_{aba}^2+p_{abb}^2=\sum_{i_1<i_2}\sum_{i_3= i_1}^{i_2}p_{i_1i_2i_3}^2\\
&\sum_{a<b}\sum_{i_1\neq a,b}p_{i_1aa}^2+p_{i_1bb}^2=(L-2)\sum_{i_1<i_2}\sum_{i_3= i_1}^{i_2}p_{i_1i_2i_3}^2\\
&\sum_{a<b}\sum_{\substack{i_1<i_2<\cdots<i_k\\i_1,i_2,\dots,i_k\neq a,b}}p_{i_1i_2\cdots i_ka}^2+p_{i_1i_2\cdots i_kb}^2=(L-k-1)\sum_{i_1<i_2<\cdots<i_k}\sum_{t\neq i_1,i_2,\dots,i_k}p_{i_1i_2\cdots i_kt}^2\\
&\sum_{a<b}\sum_{\substack{i_1<i_2<\cdots<i_k\\i_1,i_2,\dots,i_k\neq a,b}}p_{i_1i_2\cdots i_kaba}^2+p_{i_1i_2\cdots i_kabb}^2=(k+1)\sum_{i_1<i_2<\cdots<i_{k+2}}\sum_{t=i_1}^{i_{k+2}}p_{i_1i_2\cdots i_{k+2}t}^2\\
&\sum_{a<b}\sum_{\substack{i_1<i_2<\cdots<i_{k+1}\\i_1,i_2,\dots,i_{k+1}\neq a,b}}p_{i_1i_2\cdots i_{k+1}aa}^2+p_{i_1i_2\cdots i_{k+1}bb}^2=(L-k-2)\sum_{i_1<i_2<\cdots<i_{k+2}}\sum_{t=i_1}^{i_{k+2}}p_{i_1i_2\cdots i_{k+2}t}^2\\
&\sum_{a<b}\sum_{\substack{i_1<i_2<\cdots<i_{k+1}\\i_1,i_2,\dots,i_{k+1}\neq a,b}}p_{i_1i_2\cdots i_{k+1}ab}^2+p_{i_1i_2\cdots i_{k+1}ba}^2=(k+2)\sum_{i_1<i_2<\cdots<i_{k+2}}\sum_{t\neq i_1\cdots,i_{k+2}}p_{i_1i_2\cdots i_{k+2}t}^2,\\
\end{align*}
where $k=2,4,6,\dots,N$. And we define
\begin{align*}
&x_1\equiv\sum_{i_2}p_{i_1i_2}^2\\
&x_2\equiv\sum_{i_1<i_2}\sum_{i_3= i_1}^{i_2}p_{i_1i_2i_3}^2\\
&x_k\equiv\sum_{i_1<i_2<\cdots<i_k}\sum_{t=i_1}^{i_k}p_{i_1i_2\cdots i_kt}^2\\
&x_{k+1}\equiv\sum_{i_1<i_2<\cdots<i_k}\sum_{t\neq i_1,\dots,i_k}p_{i_1i_2\cdots i_kt}^2
\end{align*}
It is easy to check $\sum_{a<b}Q_{ab}=(L-1)(x_1+x_2+\cdots+x_{N+1})$. With the concavity of $\varphi(x,y)$ and Jensen’s inequality, we have 
\begin{equation}\label{eq449}
I_{AE}\le\frac{\sum_{k=1}^N\varphi[(L-k)x_k,kx_{k+1}]}{(L-1)\sum_{k=1}^{N+1}x_k},
\end{equation}
By searching the maximum value of Eq.~\eqref{eq449} in non-negative spaces,we could get the upper bound of $I_{AE}$.
Eq.~\eqref{eq49} and Eq.~\eqref{eq449} are actually the same, i.e., the expression of $I_{AE}$ is just Eq.~\eqref{eq449} whether $N$ is an odd number or even number.

\section{The simulation model of RRDPS with single photon detectors}
If Alice prepares the $L$-pulses trains with phase-randomized weak coherent state source with intensity $\mu$ and Bob's detectors are single photon detectors in RRDPS, the counting rate and error counting rate \cite{yin2018improved} are
\begin{align}
&Q_\mu=\sum_{r=1}^{L-1}\frac{L-r}{L-1}(1-p_d)^{2L-2r-1}e^{-(L-r)\eta\mu/L}(\eta\mu/L+2p_d),\\
&T_\mu=\sum_{r=1}^{L-1}\frac{L-r}{L-1}(1-p_d)^{2L-2r-1}e^{-(L-r)\eta\mu/L}p_d.
\end{align}
And the error rate is $E_\mu=T_\mu/Q_\mu$.
The final key rate of RRDPS with single photon detectors is
\begin{equation}
LR=Q_\mu(1-fH(E_\mu))-e_{src}-(Q-e_{src})\phi(n_{th},L),
\end{equation}
where $e_{src}=\sum_{k>n_{th}}\frac{\mu^ke^{-\mu}}{k!}$ and 
\begin{equation}
\phi(N,L)=\max\limits_{x_1,x_2,\dots,x_{N+1}}\left\{\frac{\sum\limits_{k=1}^N\varphi((L-k)x_k,kx_{k+1})}{L-1}\right\},
\end{equation}
and $\sum_{k=1}^{N+1}x_k=1$.

\section{The simulation models of RRDPS with yes-no detectors}
If Alice prepares the $L$-pulses trains with phase-randomized weak coherent state source with intensity $\mu$ and Bob's detectors are single photon detectors in RRDPS, the counting rate and error counting rate are
\begin{align}
\label{eq110}Q_\mu=&\sum_{r=1}^{L-1}\frac{L-r}{L-1}(1-p_d)^{2L-2r-1}e^{-\frac{\eta\mu(L-r-1)}{L}}[1-(1-p_d)e^{-\frac{\eta\mu}{L}}],\\
\label{eq111}T_\mu=&\sum_{r=1}^{L-1}\frac{L-r}{L-1}p_d(1-p_d)^{2L-2r-1}e^{-\frac{\eta\mu(L-r)}{L}}.
\end{align}

Eqs.~\eqref{eq12} and \eqref{eq53} clearly show that the bit-flip error rate of even-counts is $0.5$ no matter what Eve does. And thus the final secure key rate is 
\begin{equation}\label{finalkeyrate}
\begin{split}
&LR=\max\limits_{\gamma}\min\limits_{\alpha}\{\gamma \max[R_1(\alpha),0]+(1-\gamma) \max[R_2(\alpha),0]\},\\
&s.t. \quad 1-\frac{2EQ}{Q-e_{src}}\le \alpha \le 1,\quad 0\le\gamma\le1, \\
&R_1(\alpha)=\alpha(Q_\mu-e_{src})(1-\phi(n_{th},L))-fQ_\mu H(E_\mu),\\
&R_2(\alpha)=(1-\alpha)(Q_\mu-e_{src})-fQ_\mu H(E_\mu),
\end{split}
\end{equation}

%\bibliography{refs}

\begin{thebibliography}{22}
\bibitem{notation} Since the operation of yes-no detectors is done in the photon number space, we do not have to consider the linear superposition of different photon numbers for the incident light of the detectors.
\end{thebibliography}

\end{document}